\documentclass[reprint,aps,prl,superscriptaddress,10pt,english,nolongbibliography]{revtex4-2}

\usepackage{graphicx}
\usepackage[normalem]{ulem}
\usepackage{bm,amssymb,amsmath,mathrsfs,amstext,latexsym,physics}

\usepackage[dvipsnames]{xcolor}
\usepackage[colorlinks=true,citecolor=blue,urlcolor=blue,linkcolor=blue]{hyperref}

\newcommand{\der}[2]{\frac{d #1}{d #2}}

\begin{document}

\title{Symmetry re-breaking in an effective theory of quantum coarsening}

\noaffiliation{}

\author{Federico Balducci}
\email{fbalducci@pks.mpg.de}
\affiliation{Max Planck Institute for the Physics of Complex Systems, N\"othnitzer Str.\ 38, 01187 Dresden, Germany}

\author{Anushya Chandran}
\affiliation{Max Planck Institute for the Physics of Complex Systems, N\"othnitzer Str.\ 38, 01187 Dresden, Germany}
\affiliation{Department of Physics, Boston University, Boston, Massachusetts 02215, USA}

\author{Roderich Moessner}
\affiliation{Max Planck Institute for the Physics of Complex Systems, N\"othnitzer Str.\ 38, 01187 Dresden, Germany}

\date{\today}


\begin{abstract}
    We present a simple theory accounting for two central observations in a recent experiment on quantum coarsening and collective dynamics on a programmable quantum simulator [T. Manovitz et al., Nature \textbf{638}, 86 (2025)]: an apparent speeding up of the coarsening process as the phase transition is approached; and persistent oscillations of the order parameter after quenches within the ordered phase. Our theory, based on the Hamiltonian structure of the equations of motion in the classical limit of the quantum model, finds a speeding up already deep within the ordered phase, with subsequent slowing down as the domain wall tension vanishes upon approaching the critical line. Further, the oscillations are captured within a mean-field treatment of the order parameter field. For quenches within the ordered phase, small spatially-varying fluctuations in the initial mean-field lead to a remarkable long-time effect, wherein the system dynamically destroys its long-range order and has to coarsen to re-establish it. We term this phenomenon \emph{symmetry re-breaking}, as the resulting late-time magnetization can have a sign opposite to the initial magnetization. 
\end{abstract}

\maketitle


Understanding the out-of-equilibrium dynamics of many-body quantum systems is one of the fundamental challenges of statistical mechanics. Contrary to the equilibrium setting, where several analytical and numerical techniques are available, time evolution is difficult to solve even in comparatively simple cases. This is primarily due to the absence of perturbatively small parameters, and to the growth of entanglement, which makes it hard for classical computers to store the many-body wavefunction. A promising approach to the problem is the development of quantum simulators, i.e.\ physical platforms engineered to emulate the behavior of complex quantum systems in a highly controllable and tunable manner~\cite{Feynman1982Simulating,Lloyd1996Universal}. Unlike classical computers, quantum simulators can natively encode and evolve quantum states, allowing them to bypass many of the bottlenecks that hinder classical computation. 

Quantum simulation is expected to be particularly relevant and powerful in two-dimensional (2D) systems, where classical computational methods such as tensor networks face severe limitations~\cite{Orus2019Tensor,Banuls2023Tensor}. At the same time, physics is richer in 2D: systems can exhibit long-range order at finite temperature, and a range of topological defects~\cite{Mermin1979Topological} with complex dynamics~\cite{Pismen1999Vortices,Vinen2002Quantum,Balducci2022Localization,*Balducci2023Interface,Vodeb2024Nonequilibrium,Krinitsin2025Roughening,Pavesic2025Constrained,Trigueros2025Dynamics}. Quantum simulators, which naturally operate in higher dimensions and can directly implement 2D Hamiltonians, thus offer a unique experimental window on these complex quantum dynamical behaviors in regimes that can be otherwise hard to access. 

Our starting point is a set of recent experimental developments in the out-of-equilibrium quantum simulation of 2D many-body systems. Multiple platforms---including superconducting qubits, Rydberg atoms, and trapped ions---have probed time dynamics in 2D, in both quantum ordered and disordered phases~\cite{Andersen2025Thermalization,Manovitz2025Quantum,Haghshenas2025Digital}. Notably, one experiment~\cite{Manovitz2025Quantum} investigated quantum collective dynamics and phase-ordering kinetics (also known as coarsening), claiming deviations from classical expectations. The coarsening dynamical exponent was found to be $z_\mathrm{c}\!=\!2$, consistent with classical ``model C'' dynamics~\cite{Hohenberg1977Theory,Bray1994Theory,Cugliandolo2015Coarsening}, which is the dynamics of a non-conserved order parameter coupled to a conserved field (here the energy). However, the coarsening process accelerated as the system approached the phase transition, without exhibiting signs of critical slowing down conventionally associated with criticality. Moreover, the experiment uncovered persistent oscillations in the order parameter following quenches from deep within the ordered phase to the proximity of the critical point. This feature was linked as well to the quantum nature of the critical point~\cite{Manovitz2025Quantum}. In the remainder of this paper, we operationally refer to this set of phenomena collectively as quantum coarsening~\footnote{The term ``quantum coarsening'' was also used in studies of phase ordering in open quantum systems, where the dynamics is not unitary~\cite{Aron2009Driven,*Aron2010Coarsening}.}.

We introduce a minimal model based on the Hamiltonian dynamics of classical spins to explain the experimental observations, which suggests that the phenomena are not characteristic of quantum systems. We further develop a mean-field (MF) approximation that allows for an analytical characterization of the observed oscillations. Our analysis reveals that these oscillations can already appear deep in the ordered phase, and are clearly distinct from the long-time coarsening dynamics. The onset of the latter is determined by the exponential growth of spatial fluctuations, which drive the system into an intermediate dynamical regime where long-range order is temporarily suppressed, before being re-established through coarsening. This gives the final state a sensitive dependence on quench parameters. This sequence of events unveils a robust phenomenon, which we term \emph{symmetry re-breaking}, and should generically occur in the symmetry-broken phases of systems with Hamiltonian dynamics. We note that features of symmetry re-breaking have been previously observed in weakly perturbed fully connected Ising models~\cite{Lerose2018Chaotic,*Lerose2019Impact}.

\paragraph{Effective model of quantum coarsening.}

We consider the transverse-field Ising model, arguably the simplest two-dimensional quantum spin model exhibiting a $\mathbb{Z}_2$-symmetry-breaking transition: $\hat{H}[g] \!=\! - \sum_{\ev*{ij}} \hat{S}_i^z \hat{S}_j^z - g \sum_i \hat{S}_i^x$, where the spin-1/2 operators $\hat{S}_i$ live on a $L \times L$ square lattice, $\ev*{ij}$ indicates nearest neighbors, and we fix the exchange energy scale to 1. 

\begin{figure}[t]
    \centering
    \includegraphics[width=\columnwidth]{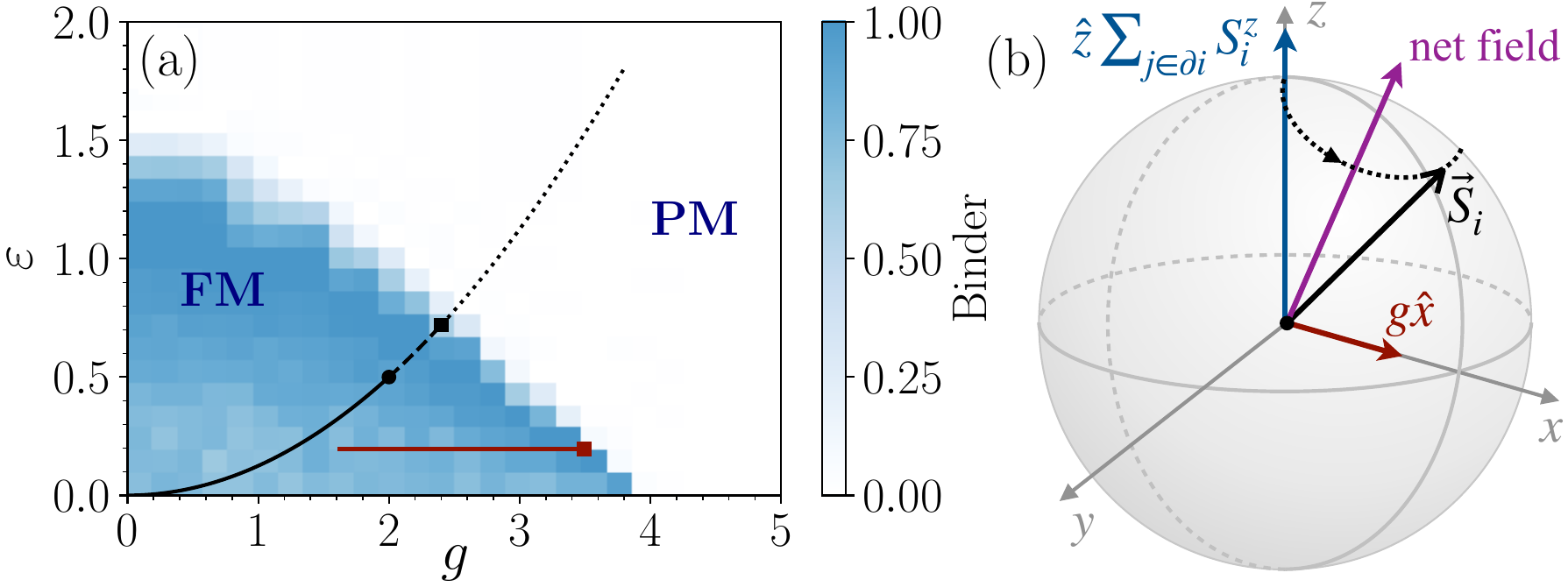}
    \caption{\emph{Phase diagram and global dynamics in the classical transverse-field Ising model}. (a) The Binder cumulant $B \!=\! 3(1 - \ev*{m^4}/3\ev*{m^2}^2)/2$ distinguishes between the ordered ($B\!=\!1$) and disordered ($B\!=\!0$) phases. The red line shows the states accessed with the areal speed simulations. The black line shows the post-quench states accessed in the oscillation experiments which exhibit either ferromagnetic (solid), paramagnetic (dotted), or symmetry re-breaking (dashed) behaviors. The square dot marks the thermodynamic transition at $g=g_c$, and the round dot the dynamical transition at $g=g_{\mathrm{dyn}}$ in the mean-field model. The data was obtained via Monte Carlo simulations at a fixed temperature, for a 50$\times$50 lattice, and plotted as a function of the excess energy density $\varepsilon$. (b) The Bloch sphere of a classical spin $i$. Switching on a transverse field misaligns the net (exchange+applied) local field with the spin direction, leading to spin precession.}
    \label{fig:phase_diag}
\end{figure}

We take the classical limit of this model by sending the spin representation from $S\!=\!1/2$ to $\infty$ while rescaling $\hat{S}_i^\alpha \to \hat{S}_i^\alpha/S$, thus obtaining the classical model~\cite{Burkhardt1974Critical}
\begin{equation}
    \label{eq:H}
    H[g] = - \sum_{\ev*{ij}} S_i^z S_j^z - g \sum_i S_i^x.
\end{equation}
The commutation relations $\big[\hat{S}_i^\alpha, \hat{S}_j^\beta\big] = i \hbar \delta_{ij} \epsilon^{\alpha\beta\gamma} \hat{S}_i^\gamma$ get mapped to the Poisson brackets $\big\{S_i^\alpha, S_j^\beta\big\} = \delta_{ij} \epsilon^{\alpha\beta\gamma} S_i^\gamma$. Therefore, the classical system is endowed with a natural Hamiltonian dynamics $\partial_t \vec{S}_i = -(g\hat{x} +\hat{z} \sum_{j \in \partial i} S_j^z) \wedge \vec{S}_i$, see Fig.~\ref{fig:phase_diag}(b), which conserves the total energy $H$. Owing to the bipartite nature of the lattice, this dynamics can be simulated to high accuracy with trotterized algorithms that conserve the energy to machine precision (see End Matter). The initial conditions of the dynamics are specified case by case.

The phase diagram of the classical model, Eq.~\eqref{eq:H}, is shown in Fig.~\ref{fig:phase_diag}(a), as a function of the transverse field $g$ and the excess energy density $\varepsilon$, which is the energy density in excess of that in the ground state. As in the quantum counterpart, a line of second-order phase transitions separates a paramagnetic (PM) phase at large $(g,\varepsilon)$ from a ferromagnetic (FM) one at small $(g,\varepsilon)$. The order parameter is the magnetization along $z$: $m = \frac{1}{L^2}\sum_i S^z_i$.

\begin{figure}
    \centering
    \includegraphics[width=\columnwidth]{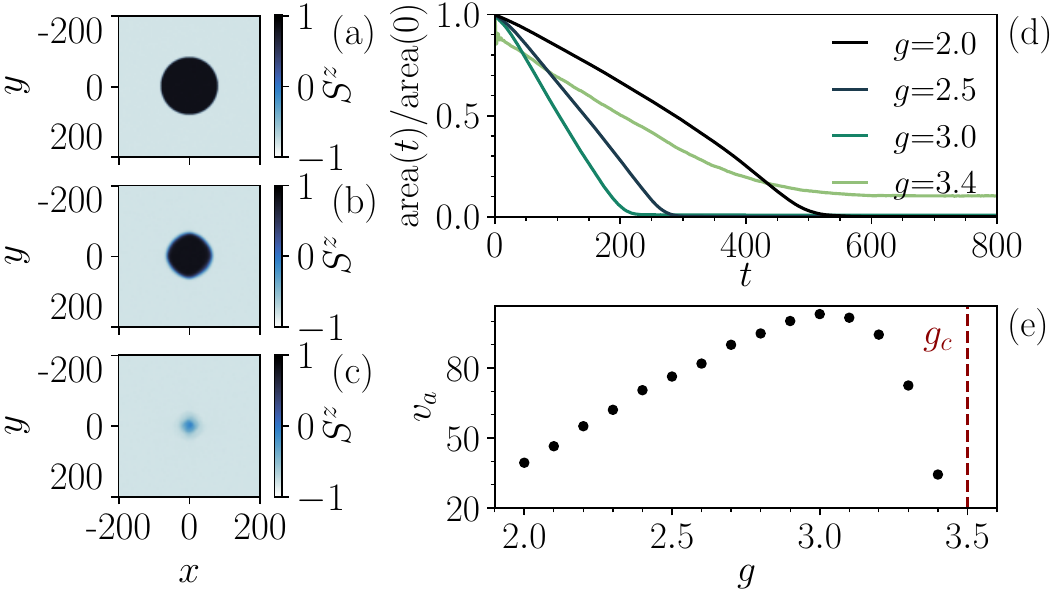}
    \caption{\emph{Areal speed of coarsening}. An initial ``up" circular domain, with the opposite magnetization as compared to its surroundings, is eroded as the system evolves with Hamiltonian dynamics [$g\!=\!2$ and times $t\!=\!0$,250,500 for panels (a),(b),(c) respectively]. (d--e) The speed $v_a$ at which this happens increases with $g$ starting from $g\!=\!0$, until it decreases in the proximity of the phase transition. The excess energy density is fixed to $\varepsilon\!=\!0.2$, as marked in Fig.~\ref{fig:phase_diag}. A lattice of size 400$\times$400 is used for the simulations.}
    \label{fig:speed}
\end{figure}

In the language of the Ising model, Eq.~\eqref{eq:H}, the experimental observations~\cite{Manovitz2025Quantum} can be stated as follows. First, suppose the system is initialized in a domain-wall configuration with one macroscopic region in the ``up'' ground state ($m\!>\!0$) and the other in the ``down'' ground state ($m\!<\!0$). The interface between such two thermodynamically-large domains spans a length $O(L)$, so the total excess energy density vanishes in the thermodynamic limit as $O(1/L)$. Thus, the initial state is close to a ground state deep in the quantum FM phase. The experiment shows that the ensuing coarsening dynamics is curvature-driven: for a circular domain of radius $r$, the radius shrinks in time according to 
\begin{equation}
    \label{eq:curvature}
    \partial_t r = -v_a/r, 
\end{equation}
where $v_a$ has the dimensions of area per unit time. The curvature-driven coarsening is compatible with the classical ``model C'' dynamics, characterized by a coarsening dynamical exponent $z_\mathrm{c}\!=\!2$~\cite{Bray1994Theory,Cugliandolo2015Coarsening}. The experiment observed a $v_a$ increasing as $g \!\nearrow\! g_c$, a feature which we analyze in detail below, see Fig.~\ref{fig:speed}.

Second, the ground state at $g_0\!=\!0$ is prepared, and then quenched to $g \! \simeq \! g_c$ on either side of the transition. The locus of the excess energy density of the post-quench states $\varepsilon \!=\! \varepsilon(g)$ is shown as a black line in Fig.~\ref{fig:phase_diag} for the classical model; a similar line can be computed for the quantum model~\cite{Blass2016Test}. After the quench, the order parameter displays underdamped oscillations in time, with a frequency $\omega$ and phenomenological damping $\gamma$. Notably, $\omega$ presents a dip in the vicinity of the phase transition line, where instead $\gamma$ is maximal, and $\omega$ (roughly) halves when crossing from the FM to the PM side. 

In the following, we address the above points, showing that they can be explained already within classical Hamiltonian dynamics. While the Ising model is different from the Rydberg Hamiltonian used in the experiments, the two can be treated with similar MF techniques and are expected to behave qualitatively alike.

\paragraph{Areal speed of coarsening.} 

To show that the classical limit $S \to \infty$ is able to capture features of the quantum spins-1/2, we investigate the (areal) speed of shrinking of a single minority domain, $v_a$, for different $(g,\varepsilon)$ in the FM phase. We take as initial state for the dynamics a circular domain, as shown in Fig.~\ref{fig:speed}(a)---a more detailed description is in the End Matter. During time evolution the disk shrinks and eventually disappears [Fig.~\ref{fig:speed}(b--c)].

Figure~\ref{fig:speed}(d) shows that the area of the domain, measured by the number of spins with $S_i^z>0$ inside the original domain boundary, decreases linearly in time, consistent with curvature-driven coarsening: $dr^2/dt \simeq \mathrm{const.} = v_a$, see also Eq.~\eqref{eq:curvature}. In Fig.~\ref{fig:speed}(e), we also show that $v_a$ increases with $g$ already well within the FM phase: this is consistent with the experimental observations. However, closer to the critical line, one can see that $v_a$ decreases again as $g$ approaches $g_c$. 

The two regimes---the areal speed increasing and then decreasing---reflect different physical mechanisms. Deep within the FM phase, $v_a$ increases with $g$ because $g$ defines the strength of the ``kinetic term'' $-g \sum_i S_i^x$: making it larger speeds up the spin dynamics generally, and in particular at the domain wall edges. In the quantum model, this heuristic can be made rigorous with prethermalization bounds as $g \to 0$~\cite{Abanin2015Exponentially,Mori2016Rigorous,Abanin2017Rigorous,Balducci2022Localization,*Balducci2023Interface}, but similar arguments can apply to classical models~\cite{DeRoeck2025Long}. Separately, as $g$ approaches $g_c$, the vanishing of the domain-wall tension provides a mechanism for decreasing the areal speed $v_a$, as is evident in Fig.~\ref{fig:speed}(e).

\paragraph{Post-quench oscillations.}

We now study quench dynamics within the ordered phase. To compare with experiments, we prepare the system at $g\!=\!0$ and fixed excess energy density $\varepsilon_0 \gtrsim 0$. A nonzero excess energy density is essential, as the classical model's ground state is uniform and, in the absence of fluctuations, the uniform (zero-momentum) mode decouples from the dynamics. We perform a sudden quench by evolving the system with the Hamiltonian $H[g]$ for $g\!>\!0$. The locus of the excess energy density of the post-quench states is marked with the black line in Fig.~\ref{fig:phase_diag} for the case $\varepsilon_0 \!=\! 0$; the lines corresponding to $\varepsilon_0 \! \gtrsim\! 0$ lie above the displayed one.

\begin{figure}
    \centering
    \includegraphics[width=\columnwidth]{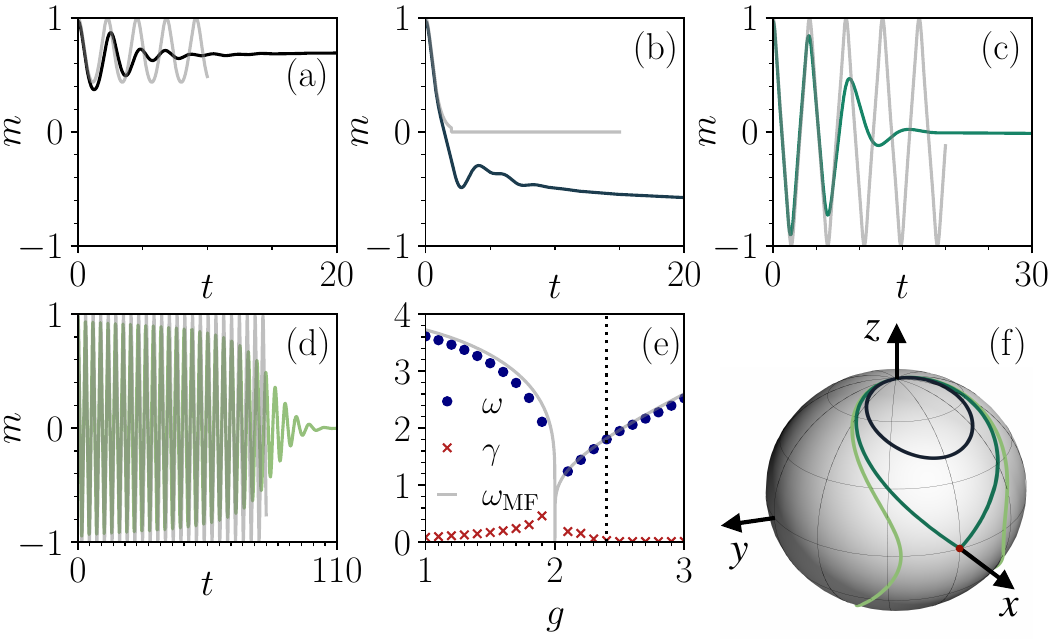}
    \caption{\emph{Post-quench oscillations}. (a--d) After the quench, the magnetization (colored lines) oscillates at the frequency set by the mean-field solution (pale grey lines), while getting damped. The plots are respectively for $g=1.8,2.0,2.2,2.5$; the thermodynamic phase transition is at $g_c \simeq 2.4$. (e) Comparison of the oscillation frequencies with the mean-field prediction in gray, showing excellent agreement. The frequency minimum does {\it not} correspond to the Ising transition (vertical dotted line). Notice the absence of damping $\gamma$ in the mean field theory. (f) Bloch sphere representation of each spin in the uniform state, showing a trajectory confined to one half of the Bloch sphere, one crossing to the other half of the sphere, and the separatrix between the two. A lattice of size 200$\times$200 is used for the simulations.}
    \label{fig:zero_mode}
\end{figure}

\begin{figure}
    \centering
    \includegraphics[width=0.9\columnwidth]{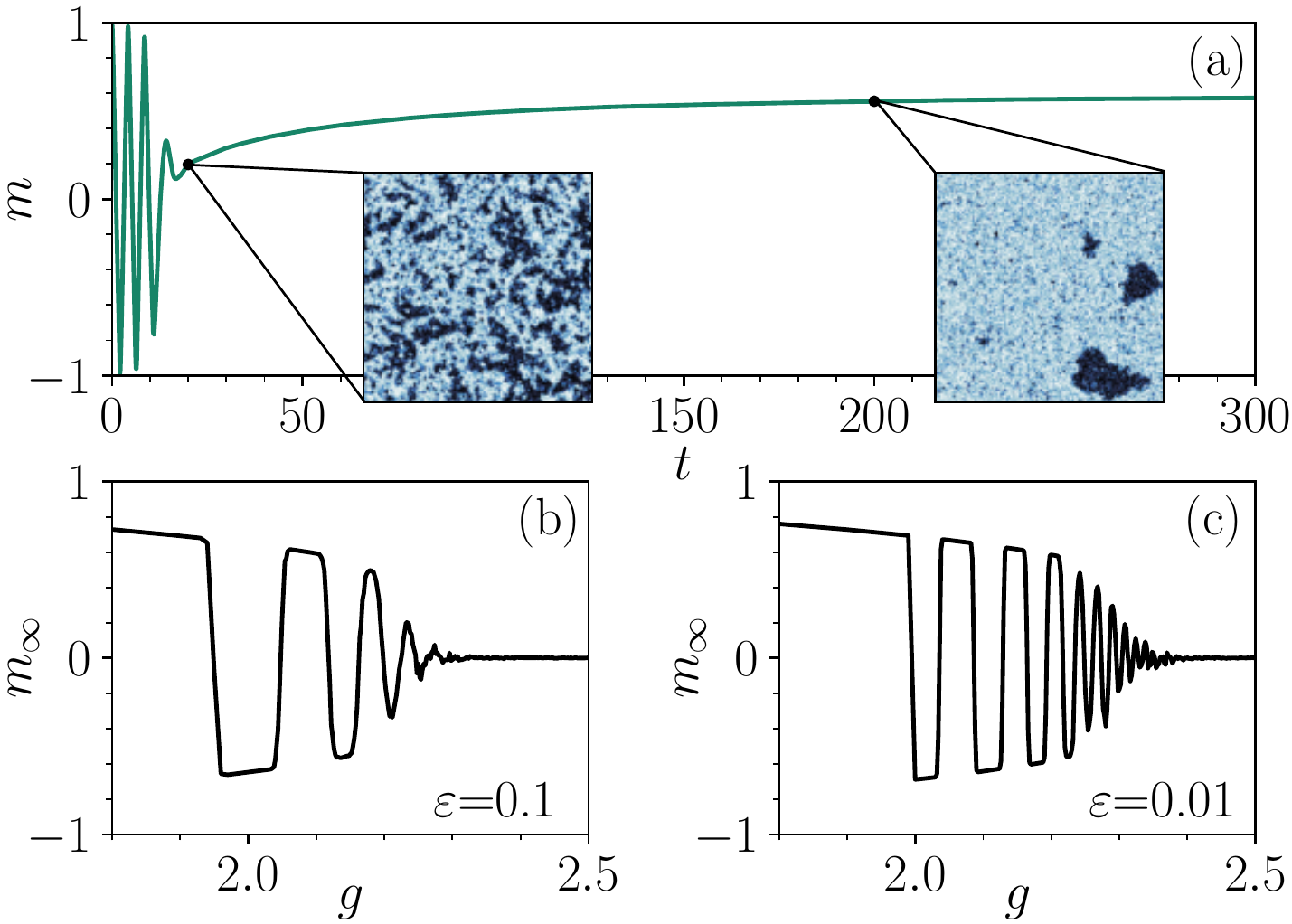}
    \caption{\emph{Symmetry re-breaking}. (a) After the magnetization oscillations die out, the snapshot shows that the system is locally ordered, but globally broken into domains (colorbar identical to Fig.~\ref{fig:speed}). These domains coarsen in time, eventually leading to the reestablishment of a global magnetization at one of its two equilibrium values. (b--c) The sign of the steady-state magnetization $m_\infty$ is determined by the parity of the number of oscillations the system undergoes before they are damped out. This number increases with decreasing energy density of the pre-quench state. A lattice of size 200$\times$200 is used for the simulations.}
    \label{fig:rebreak}
\end{figure}

Figure~\ref{fig:zero_mode} shows that the quench protocol induces large-scale oscillations of the order parameter $m$, at $\varepsilon_0=0.2$. Deep in the FM phase, these oscillations remain confined to $m\!>\!0$, indicating the persistence of the initial symmetry-broken state (Fig.~\ref{fig:zero_mode}(a)), while $m$ takes both signs in the PM, eventually relaxing to zero (Fig.~\ref{fig:zero_mode}(d)). However, there is a finite region within the FM in which $m$ changes sign, either once (Fig.~\ref{fig:zero_mode}(b)) or several times (Fig.~\ref{fig:zero_mode}(c)) before relaxing to a value which slowly drifts with time towards the equilibrium value (Fig.~\ref{fig:rebreak}(a)). 

These features can be understood in a MF analysis of the post-quench dynamics. The first step is to notice that, when a spatially uniform configuration $\vec{S}_i \equiv \vec{S}$ is evolved with the classical Hamiltonian, the zero-momentum mode decouples completely. The effective equations of motion describing its dynamics are, self-consistently, of MF type:
\begin{equation}
    \label{eq:LMG}
    \partial_t \vec{S} = -\big(4 \hat{z} S^z + g \hat{x} \big) \wedge \vec{S},
\end{equation}
where the factor of 4 reflects the coordination number of the square lattice. These equations of motion correspond to those of the fully-connected, Lipkin-Meshkov-Glick (LMG) Hamiltonian $H= -2 (S^z)^2 - g S^x$~\cite{Meshkov1965Validity}. Since the phase space of the MF model is two-dimensional (i.e.\ the surface of the unit sphere) and there is one conserved quantity (the energy), the dynamics is regular and the spin precesses along closed loops, see Fig.~\ref{fig:zero_mode}(f). The LMG oscillation frequency $\omega(g)$, computed in the  End Matter, agrees quantitatively with the frequency extracted from the oscillations in the 2D model, see Fig.~\ref{fig:zero_mode}(e). Observe that the minimum of the oscillation frequency at $g\!=\!2$ does \emph{not} occur at the thermodynamic phase transition of the 2D model ($g_c \simeq 2.4$ for the energy density used). Instead, a \emph{dynamical} phase transition occurs at $g\!=\!g_\mathrm{dyn}\!=\!2$ in the LMG model~\cite{Zunkovic2016Dynamical,Lerose2018Chaotic,*Lerose2019Impact}, manifesting in the Bloch sphere as a separatrix (Fig.~\ref{fig:zero_mode}(f)). To one side of the separatrix ($g\!<\!2$), the magnetization is confined to the upper-half sphere $S^z\!>\!0$, while on the other side ($g\!>\!2$) $S^z$ takes both signs. This is also reflected in the maximum of the phenomenological damping rate $\gamma$---as obtained by fitting the underdamped oscillations---being at $g\!=\!2$, see Fig.~\ref{fig:zero_mode}(e).

\paragraph{Symmetry re-breaking.}

We are now in position to explain quenches in the FM phase, in which oscillations drive the magnetization to near zero, or even reverse its sign (Figs.~\ref{fig:zero_mode}(b--c)) at late times. Observe that in the LMG model, the quench to $g\!=\!2$ (Fig.~\ref{fig:zero_mode}(b)) corresponds to a separatrix trajectory that terminates at the unstable PM fixed point $\vec{S} = \hat{x}$. In practice, the LMG description holds locally for the 2D model, and small fluctuations shift the local MF to either one or the other side of the separatrix. The result is that some regions of the 2D lattice end up ``bouncing back'' to $S_i^z>0$, while others cross over to $S_i^z<0$. As the order set by the initial condition is lost, self-consistently the MF description ceases to be valid. As the system is in the FM phase in equilibrium, it coarsens again {\it without ever leaving the ordered phase}, to re-establish the magnitude of the magnetization in thermodynamic equilibrium. We term this process \emph{symmetry re-breaking}.

Symmetry re-breaking takes place in the whole region $g_\mathrm{dyn} \!\leq\! g \!<\! g_c$. For $g\!>\!g_\mathrm{dyn}\!=\!2$, the magnetization performs a few oscillations. The oscillations feature a non-exponential relaxation, which terminates consistently at the same time across different realizations of the initial (noisy) state, and at a non-thermal value for $m$. We investigate this feature in detail in the next section. Figure~\ref{fig:rebreak}(a) shows that, at the end of the oscillations, the system is partitioned into different domains of ``up'' and ``down'' spins, the relative abundance of which seems fixed to a certain ratio. These domains represent the seed for coarsening dynamics, which drives the formation of larger and larger domains. Due to the imbalance of the number of ``up'' and ``down'' spins in the post-oscillation states, the system ends with either a positive or negative magnetization $m_\infty$ in the infinite-time limit, as shown in Fig.~\ref{fig:rebreak}(b--c). The dependence of the sign of $m_\infty$ on the quench parameters $g,\varepsilon$ seems highly complicated, while its absolute value is set by the microcanonical ensemble.

\paragraph{Role of fluctuations.}

Symmetry re-breaking is a mechanism driven by an interplay of a MF and spatially-varying fluctuations on top of it. The behavior of the MF is fully captured by Eq.~\eqref{eq:LMG}; here we analyze the fluctuations by developing a self-consistent gaussian theory thereof. Doing so for the $S=\infty$ Ising model results in a complicated set of equations, that are difficult to analyze~\cite{Lerose2019Impact}. For this reason, we investigate the same phenomenon in a simpler $\phi^4$ theory in 2D. As a side result, this showcases the generality of our findings beyond the specific choice of model.

The starting point is the 2D $\phi^4$ theory with Hamiltonian
\begin{equation}
    \label{eq:phi4}
    H = \sum_i \left[ \frac{\pi_i^2}{2} + \frac{(\nabla \phi)_i^2}{2} + \frac{r}{2} \phi_i^2 + \frac{u}{4} \phi_i^4 \right],
\end{equation}
where $\nabla$ represents the lattice derivative: $(\nabla \phi)_{(x,y)} \equiv (\phi_{(x+1,y)}-\phi_{(x,y)}, \phi_{(x,y+1)}-\phi_{(x,y)})$ for $i\!=\!(x,y)$, $\phi_i$ is a scalar field, and $\pi_i$ the conjugate momentum. The presence of the conjugate momenta $\pi_i$ ensures that the dynamics is Hamiltonian. As described in the End Matter, we single out the MF by passing to Fourier space $\phi_i \to \tilde{\phi}_k$: the MF corresponds to the $k\!=\!0$ mode, while the $k \neq 0$ modes encode spatial variations in the scalar field. Then, we approximate the $k \neq 0$ modes as gaussian fluctuations on top of the MF by replacing $\tilde{\phi}^4 \to 3 \ev*{\tilde{\phi}^2} \tilde{\phi}^2$ using Wick's theorem. After these approximations, the equations of motion become
\begin{subequations}
    \label{eq:gaussian}
    \begin{align}
        \label{eq:gaussian_MF}
        \partial^2_t\tilde{\phi}_0 &= -r\tilde{\phi}_0 - u\tilde{\phi}_0 \left(\tilde{\phi}_0^2 + 3\ev*{\tilde{\phi}^2}_c \right), \\
        \label{eq:gaussian_fluct}    
        \partial^2_t\tilde{\phi}_k &=-\left[\epsilon_k + r +3u \left( \tilde{\phi}_0^2 + \! \ev*{\tilde{\phi}^2}_c \right) \right] \tilde{\phi}_k, \quad k \neq 0,
    \end{align}
\end{subequations}
where $\epsilon_k \!=\! 4-2\cos k_x -2 \cos k_y$. These are the equations of motion of coupled oscillators, one of which is nonlinear (i.e.\ $\tilde{\phi}_0$), while the others are coupled to each other only through $\ev*{\tilde{\phi}^2}_c \equiv \frac{1}{N} \sum_{k\neq 0} \tilde{\phi}_k \tilde{\phi}_{-k}$.
 
\begin{figure}
    \centering
    \includegraphics[width=\columnwidth]{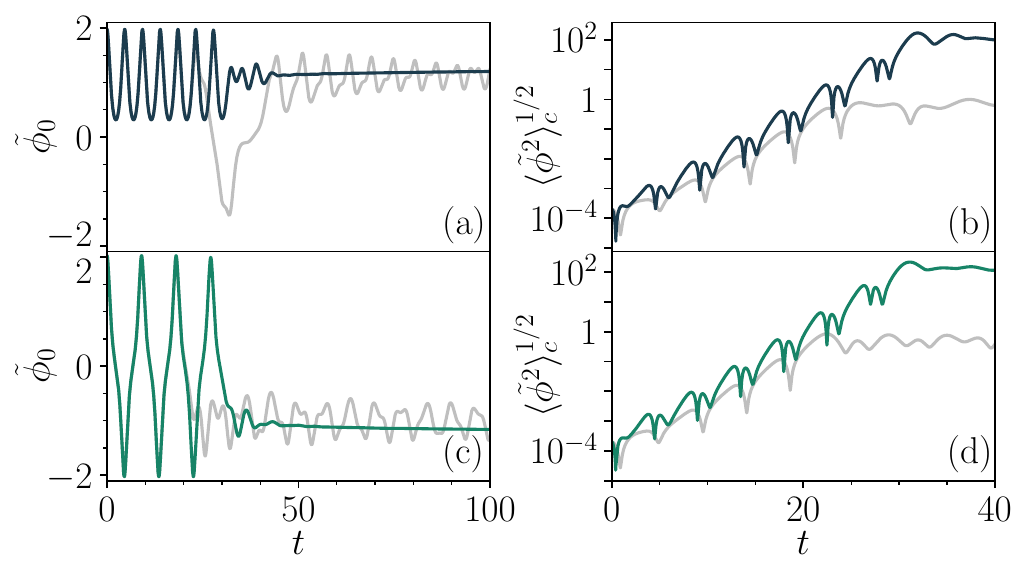}
    \caption{\emph{Role of fluctuations}. Order parameter oscillations (a) and growth of fluctuations (b) for the 2D $\phi^4$ theory, Eq.~\eqref{eq:phi4}, for a quench $r_i\!=\!-3.9$ to $r\!=\!-2$. The exact dynamics (solid blue lines) agrees with the gaussian decoupling (Eq.~\eqref{eq:gaussian}, pale gray lines) until the fluctuations become of the same order of the mean field. Panels (c) and (d) similarly show a quench into the symmetry re-breaking region ($r_i\!=\!-4.1$, $r\!=\!-2$), for a case where the asymptotic value of the order parameter has the opposite sign. A lattice of size 201$\times$201 is used for the simulations.}
    \label{fig:fluct}
\end{figure}

In Fig.~\ref{fig:fluct}(c), we show the solution of Eqs.~\eqref{eq:gaussian}, alongside the solution of the full 2D $\phi^4$ model, for a quench protocol equivalent to that studied for 2D Ising model. One can see that both the gaussian model, Eqs.~\eqref{eq:gaussian}, and the interacting model, Eq.~\eqref{eq:phi4}, display short-time oscillations of the order parameter, akin to those observed for the classical Ising model. At the same time, the fluctuations grow exponentially in time: when the latter become of the same order of magnitude of the MF itself, the MF oscillations are disrupted. 

We trace the exponential growth of fluctuations to the presence of an ``inverted mass'' term. Namely, Eq.~\eqref{eq:gaussian_fluct} for low momenta acquires the form $\partial_t^2 \tilde{\phi}_k \approx - (r + 3u \, \tilde{\phi}_0^2) \tilde{\phi}_k$: since $\tilde{\phi}_0^2$ passes through 0 and $r\!<\!0$, the equation takes periodically the form of an inverted oscillator, making fluctuations grow exponentially in time. The exponential growth of fluctuations explains why the observed decay of the MF mode is not exponential: the oscillations are rapidly cut off when the fluctuations become of the same order as the value of the MF, but not before. One can finally notice that the full interacting model and the gaussian approximation behave differently after the oscillations have decayed. In fact, the 2D interacting model undergoes energy-conserving coarsening dynamics, as shown for the classical Ising model, while the gaussian model remains non-thermal up to infinite time~\cite{Chandran2013Equilibration,Maraga2015Aging,Chiocchetta2017Dynamical,Giachetti2025Universality}.

\paragraph{Outlook.}

We have provided a transparent explanation of two notable experimental observations in a quantum coarsening experiment~\cite{Manovitz2025Quantum}. First, we have argued that the observed speed-up is explained by the tuning parameter playing a dual role, i.e.\ tuning the system closer to the phase transition but also tuning the strength of the kinetic term. While our argument was heuristic, it can be made more rigorous through a Schrieffer-Wolff-like transformation. Second, we have thoroughly characterized the observed order parameter oscillations, predicting their shape and persistence. In doing so we have unveiled the generic phenomenon of symmetry re-breaking, in which quenches {\it entirely within} an ordered phase can lead to the destruction of order at intermediate times, and its reemergence with either sign at asymptotically long times. Our results reveal that, even in dimension as low as $d\!=\!2$, and for the ``most quantum'' spins-1/2, semiclassical analyses are a viable tool for disentangling purely quantum effects from other many-body phenomena. Further studies are clearly called for, on any of the existing platforms, in particular to investigate symmetry re-breaking, and the detailed dynamics in the critical region. A first, viable analysis is the use of a $1/S$ expansion around the classical trajectories studied in this work.

\acknowledgments
It is a pleasure to acknowledge discussion with Andrea Gambassi, Markus Heyl, Johannes Hofmann, David Huse, Jorge Kurchan, Giuseppe Mussardo and Carlo Vanoni. This work was in part supported by the Deutsche Forschungsgemeinschaft under grants FOR 5522 (project-id 499180199) and the cluster of excellence ct.qmat (EXC 2147, project-id 390858490).

\bibliography{references}

\appendix

\begin{center}
    \textbf{END MATTER}
\end{center}
\vspace{-1cm}

\section{Algorithms for the numerical simulations}

In this work, we consider two classical lattice models. The first one is the classical Ising model in transverse field [Eq.~\eqref{eq:H} in the main text]
\begin{equation}
    H[g] = - \sum_{\ev*{ij}} S_i^z S_j^z - g \sum_i S_i^x.
\end{equation}
Its Hamilton equations of motion are
\begin{equation}
    \partial_t \vec{S} = \bigg(-g\hat{x} - \hat{z} \sum_{j \in \partial i} S_j^z \bigg) \wedge \vec{S}.
\end{equation}
On bipartite lattices, like the square lattice employed in the main text, these equations can be integrated with the energy conserved up to machine precision. This is done by trotterizing the time evolution: first, the spins in sublattice $A$ are evolved for a time $dt/2$ using the magnetic field generated by the spins in sublattice $B$; then the spins in sublattice $B$ are evolved for the same time $dt/2$ keeping fixed the spins in sublattice $A$. For $dt$ small enough, the flow generated in phase space is Hamiltonian. In practice, we do not just use a very small $dt$, but also a higher-order trotterization (with 6 $A/B$ alternations per time-step)~\cite{Tsai2004Symplectic}.

The second model considered is a two-dimensional $\phi^4$ theory on the square lattice [Eq.~\eqref{eq:phi4} in the main text]
\begin{equation}
    H = \sum_i \left[ \frac{\pi_i^2}{2} + \frac{(\nabla \phi)_i^2}{2} + \frac{r}{2} \phi_i^2 + \frac{u}{4} \phi_i^4 \right],
\end{equation}
where $\nabla$ is the lattice derivative: $(\nabla \phi)_{(x,y)} \equiv (\phi_{(x+1,y)}-\phi_{(x,y)}, \phi_{(x,y+1)}-\phi_{(x,y)})$ for $i\!=\!(x,y)$. The Hamilton equations of motion are
\begin{subequations}
\begin{align}
    \partial_t \phi_{(x,y)} &= \pi_{(x,y)}, \\
    \partial_t \pi_{(x,y)} &= \phi_{(x+1,y)} + \phi_{(x-1,y)} + \phi_{(x,y+1)} + \phi_{(x,y-1)} \nonumber \\
    &\phantom{==}- 4\phi_{(x,y)} - r\phi_{(x,y)} - u\phi_{(x,y)}^3.
\end{align}
\end{subequations}
These were integrated with a standard velocity-Verlet algorithm.

\section{Initial conditions for the numerical simulations}

For the study of the areal velocity, we take as initial state for the dynamics a circular domain of radius $r(t\!=\!0)\!=\!80$ lattice sites, as in Fig.~\ref{fig:speed}(a). The transverse field $g$ is fixed through the Hamiltonian parameters. To fix $\varepsilon$, we first let the outside state be the ``down'' ground state ($S_i^x\!=\!g/4$, $S_i^z \!=\! - \sqrt{1-g^2/16}$), and the one inside the disk to be the ``up'' ($S_i^x\!=\!g/4$, $S_i^z \!=\! + \sqrt{1-g^2/16}$). We then slowly heat up the configuration by coupling it to an effective bath (via Monte Carlo sampling), until it reaches the target excess energy density $\varepsilon$. Notice that the resulting configurations are not drawn from the Gibbs distribution, as we do not run the Monte Carlo algorithm for long enough. We finally time-evolve the resulting state.

For the study of the post-quench oscillations, we need to prepare the system at $g\!=\!0$ and fixed excess energy density $\varepsilon_0 \gtrsim 0$. Again, a nonzero excess energy density is obtained by warming up gradually the perfect ferromagnet $\vec{S}_i \!=\! + \hat{z}$ via Monte Carlo sampling.

\section{Mean-field frequency of oscillation}

We show here how one can obtain the frequency $\omega(g)$ from Eq.~\eqref{eq:LMG} [main text]. The starting point is to realize that the mean-field, Lipkin-Meshkov-Glick Hamiltonian
\begin{equation}
    H_\mathrm{LMG} = -2 (S^z)^2 - g S^x
\end{equation}
defines a classical Hamiltonian system. Canonically conjugated variables for the system can be taken to be $q,p$ such that
\begin{equation}
    S^x = p, \quad
    S^y = \sqrt{1-p^2} \sin q, \quad
    S^z = \sqrt{1-p^2} \cos q,
\end{equation}
with $\{q,p\}=1$ and $q\in [-\pi,\pi]$, $p\in [-1,1]$. In these new variables the Hamiltonian reads
\begin{equation}
    \label{eq:LMG_pq}
    H_\mathrm{LMG} = -2(1-p^2) \cos^2 q - gp.
\end{equation}
One can now obtain the period of the trajectory at energy $E$ through the formula
\begin{equation}
    \label{eq:period}
    T(E) = \der{}{E} \oint p \,dq. 
\end{equation}
Notice that the energy must in the end be fixed to $E=-2$, as the ferromagnetic initial condition corresponds to $q=0$, $p=0$. The curve $p(q;E)$ at energy $E$ is obtained by solving Eq.~\eqref{eq:LMG_pq} wrt $p$:
\begin{equation}
    p_\pm(q;E) = \frac{g \pm \sqrt{g^2 + 8E\cos^2 q + 16 \cos^4 q}}{4 \cos^2 q}.
\end{equation}
The integral in Eq.~\eqref{eq:period} can be now expressed in terms of elliptic functions, but we found more convenient to just bring the derivative under the integral sign, and perform the integration numerically, finally obtaining $\omega = 2\pi/T$ as displayed in Fig.~\ref{fig:zero_mode}(e) [main text].

\section{Gaussian decoupling of the $\phi^4$ theory}

We explain here how to perform the gaussian decoupling for the 2D $\phi^4$ theory, by separating the mean-field and the fluctuations around it. The real-space Hamiltonian is [Eq.~\eqref{eq:phi4} in the main text]
\begin{equation}
    H = \sum_i \left[ \frac{\pi_i^2}{2} + \frac{(\nabla \phi)_i^2}{2} + \frac{r}{2} \phi_i^2 + \frac{u}{4} \phi_i^4 \right].
\end{equation}
Introducing the Fourier modes
\begin{equation}
    \tilde{\phi}_k = \sum_r e^{-ikr} \phi_r, \qquad
    \tilde{\pi}_k = \sum_r e^{-ikr} \pi_r,
\end{equation}
the Hamiltonian becomes
\begin{multline}
    H = \frac{1}{N} \sum_k \left[ \frac{\tilde{\pi}_k \tilde{\pi}_{-k}}{2} + \frac{\epsilon_k + r}{2} \tilde{\phi}_k\tilde{\phi}_{-k} \right] \\
    + \frac{u}{4N^3} \sum_{k_1k_2k_3} \tilde{\phi}_{k_1} \tilde{\phi}_{k_2} \tilde{\phi}_{k_3} \tilde{\phi}_{-k_1-k_2-k_3},
\end{multline}
where $N\!=\!L^2$ is the number of spins on a square lattice of side $L$.

In the situation considered in the main text, the zero mode $\tilde{\phi}_0$ is initially of $O(N)$, while the fluctuations $\tilde{\phi}_k$ are very small. Therefore, we rescale $\tilde{\phi}_0 \to N\tilde{\phi}_0$, making the zero mode intensive for convenience, and single it out in the Hamiltonian:
\begin{multline}
    H = N \left[\frac{\tilde{\pi}_0^2}{2} + \frac{r}{2} \tilde{\phi}_0^2 + \frac{u}{4} \tilde{\phi}_0^4 \right] \\
    + \frac{1}{N} \sum_{k\neq 0} \left[ \frac{\tilde{\pi}_k \tilde{\pi}_{-k}}{2} + \frac{\epsilon_k + r + 3u\tilde{\phi}_0^2}{2} \tilde{\phi}_k\tilde{\phi}_{-k} \right] \\
    + \frac{u}{4N^3} {\sum_{k_1k_2k_3}}' \tilde{\phi}_{k_1} \tilde{\phi}_{k_2} \tilde{\phi}_{k_3} \tilde{\phi}_{-k_1-k_2-k_3},
\end{multline}
where the prime on the sum indicates that the terms containing $\tilde{\phi}_0$ and $\tilde{\phi}_0^2$ were removed. Finally, we enforce gaussianity of the fluctuations by substituting $\tilde{\phi}_k^4$ with $3 \ev*{\tilde{\phi}_k^2} \tilde{\phi}_k^2$ (Wick's theorem), where the average $\ev*{\cdots}$ refers to an \emph{ensemble} average over gaussian states: 
\begin{equation}
    \ev*{\tilde{\phi}^2}_c \equiv \mathbb{E}_\mathrm{gaus} \frac{1}{N} \sum_{k \neq 0} \tilde{\phi}_k \tilde{\phi}_{-k}.
\end{equation}
Thanks to this ensemble average, the fluctuating modes $\tilde{\phi}_k$ become effectively quadratic. It then holds
\begin{multline}
    H= N \left[\frac{\tilde{\pi}_0^2}{2} + \frac{r}{2} \tilde{\phi}_0^2 + \frac{u}{4} \tilde{\phi}_0^4 \right] \\
    + \frac{1}{N} \sum_{k\neq 0} \left[ \frac{\tilde{\pi}_k \tilde{\pi}_{-k}}{2} + \frac{\epsilon_k + r + 3u(\tilde{\phi}_0^2 + \ev*{\tilde{\phi}^2}_c)}{2} \tilde{\phi}_k\tilde{\phi}_{-k} \right].
\end{multline}
One can check that the Hamilton equations of motion become 
\begin{subequations}
\begin{align}
    \partial_t \tilde{\phi}_0 &= \tilde{\pi}_0, \\
    \partial_t\tilde{\pi}_0 &= -r\tilde{\phi}_0 - u\tilde{\phi}_0 \left(\tilde{\phi}_0^2 + 3\ev*{\tilde{\phi}^2}_c \right), \\
    \partial_t \tilde{\phi}_k &= \tilde{\pi}_k, \\
    \partial_t\tilde{\pi}_k &=-\left[\epsilon_k + r +3u \left( \tilde{\phi}_0^2 + \ev*{\tilde{\phi}^2}_c \right) \right] \tilde{\phi}_k,
\end{align}
\end{subequations}
where $k$ is assumed to be nonzero. These coincide with Eqs.~\eqref{eq:gaussian} in the main text. 

The ensemble of gaussian states is fully characterized by its second moments: $\ev*{\tilde{\pi}_k \tilde{\pi}_{-k}}$, $\ev*{\tilde{\pi}_k \tilde{\phi}_k}$ and $\ev*{\tilde{\phi}_k \tilde{\phi}_{-k}}$. The evolution of these in time can be obtained by means of the equations
\begin{subequations}
\label{eq:variances}
\begin{align}
    \partial_t \ev*{\tilde{\pi}_k \tilde{\pi}_{-k}} &= -2 \omega^2_k \ev*{\tilde{\pi}_k \tilde{\phi}_k}, \\
    \partial_t \ev*{\tilde{\pi}_k \tilde{\phi}_k} &= \ev*{\tilde{\pi}_k \tilde{\pi}_{-k}} - \omega^2_k \ev*{\tilde{\phi}_k \tilde{\phi}_{-k}},\\
    \partial_t \ev*{\tilde{\phi}_k \tilde{\phi}_{-k}} &= 2\ev*{\tilde{\pi}_k \tilde{\phi}_k},
\end{align}
with
\begin{equation}
    \omega_k^2 = \epsilon_k + r + 3u\left( \tilde{\phi}_0^2 + \ev*{\tilde{\phi}_k \tilde{\phi}_{-k}} \right),
\end{equation}
that need to be evolved together with the mean-field:
\begin{equation}
    \partial_t^2\tilde{\phi}_0 = -r\tilde{\phi}_0 - u\tilde{\phi}_0 \left(\tilde{\phi}_0^2 + 3\ev*{\tilde{\phi}_k \tilde{\phi}_{-k}} \right).
\end{equation}
\end{subequations}
Equations~\eqref{eq:variances} are the ones that we simulated numerically, with a 4th order Runge-Kutta algorithm.

\end{document}